\documentclass[10pt]{article}
\usepackage[paper=letterpaper,hmargin=1.7cm,vmargin=1.7cm,footskip=.9cm]{geometry}
\usepackage{caption}
\usepackage{xurl}
\usepackage{tabularray}
\usepackage{amsmath}
\usepackage{amssymb}
\usepackage{algorithm}
\usepackage[commentColor=black,beginComment=//~,endComment={},beginLComment=//~,endLComment={},italicComments=false]{algpseudocodex}
\tikzset{algpxIndentLine/.style={draw=black}}
\algrenewcommand{\alglinenumber}[1]{\bfseries\footnotesize #1}
\algrenewcommand{\textproc}{}
\algrenewcommand{\algorithmicrequire}{\textbf{Input:}}
\algrenewcommand{\algorithmicensure}{\textbf{Output:}}
\usepackage[caption=false,font=footnotesize]{subfig}
\title{WorldMark: A Plug-and-Play World Knowledge Interface for Cross-Host Language Model Watermarking}
\author{Xiao Song, Yuan Yuqi, Zhang Yanshuo, Zhang Kejun}
\begin{document}
\raggedbottom
\maketitle
\begin{abstract}
Watermarking traces the provenance of text produced by large language models by embedding statistically detectable signals during decoding. Existing schemes fall into logits-based, sampling-based, entropy-aware, and adaptive-strength families, yet all of them place watermark signals according to local token statistics. In the open-ended text-generation settings evaluated in this work, local statistics may provide insufficient guidance for placing robust watermark signals. We introduce WorldMark, a plug-and-play interface that uses World Knowledge Memory (WKM) to organize semantic and episodic knowledge in a memory graph, converts the retrieved knowledge into a token-level knowledge saliency score, and adjusts the strength of a host watermark through Asymmetric Knowledge Modulation (AKM). WorldMark requires no backbone retraining and introduces no additional detector-side model or parameter. On the primary C4 evaluation, the complete WorldMark interface improves clean and attacked detection across three adaptive-strength host variants while slightly reducing perplexity. Additional pilot experiments on C4 and OpenGen show that direct memory conditioning transfers across multiple watermark families but can be unstable without saliency-aware modulation. WorldMark requires no additional detector-side model or parameter and introduces negligible overhead under the primary protocol.
\end{abstract}
\section{Introduction}
Large language models now generate text at a scale that makes reliable attribution a prerequisite for accountability, and watermarking has become the leading mechanism for this purpose. A watermark embeds an imperceptible yet algorithmically detectable pattern into generated text so that a verifier holding a secret key can later decide whether a passage originated from a watermarked model. Given a prompt and previously generated tokens, a watermarking method modifies either the logits or the sampling step of the decoder so that the resulting sequence carries a signal that a hypothesis test can separate from unwatermarked text at a controlled false positive rate.
\par
Existing methods divide into four families: logits based methods that bias a green list of the vocabulary \cite{abs-2301-10226,abs-2306-17439}, sampling based methods that steer the selection step through a pseudo random sequence \cite{aaronson2022watermark,abs-2307-15593}, entropy aware methods that restrict watermarking to high entropy positions to protect low entropy text such as source code \cite{abs-2305-15060,abs-2403-13485,abs-2505-14112}, and adaptive strength methods that tune the intensity per token against a text quality objective \cite{abs-2505-11541}. Despite their differences, the four families share one assumption: the decision of where and how strongly to watermark a token can be made from local information, either the token logits or a short prefix hash. This assumption may become particularly problematic in knowledge-intensive settings, where load-bearing tokens are entities, relations, and identifiers whose correctness is determined by external knowledge rather than by local uncertainty. Local statistics either mark such tokens too weakly, since they appear low entropy, or place the signal where a paraphrase attack can remove it, since the placement ignores which tokens are semantically anchored. What is missing is a placement signal that reflects which tokens carry confirmed world knowledge. We refer to this signal as knowledge saliency.
\par
\textbf{World Knowledge Memory.} World Knowledge Memory (WKM), which organizes semantic and episodic knowledge in a structured memory graph, provides the information required to construct this missing placement signal. Within WKM, semantic memory stores factual triplets extracted from the context, whereas episodic memory records the observations in which those triplets were confirmed \cite{abs-2407-04363}. By identifying which tokens instantiate confirmed facts, such a model marks tokens that are both safe to carry a watermark and hard for an attacker to rewrite without changing meaning. The remaining challenge is to convert the retrieved structured knowledge into a knowledge saliency score that can be shared across different host-watermark families without backbone retraining or any additional detector-side model or parameter.
\par
We address this challenge with WorldMark, a plug-and-play interface between World Knowledge Memory and a host watermark. WorldMark uses WKM to construct a memory graph from the prompt and generated context, retrieves the semantic triplets and episodic vertices relevant to the current generation step, and converts the retrieved knowledge into a knowledge saliency score $s_t$. The score is then used by Asymmetric Knowledge Modulation to adjust the strength of the host watermark. In the experiments reported in this paper, the complete WorldMark interface is evaluated primarily on C4 using three adaptive-strength MorphMark variants. A separate small-scale pilot on C4 and OpenGen examines direct prompt-level memory conditioning across several host watermark families. The primary evaluation shows consistent improvements over the reproduced MorphMark baselines, whereas the pilot reveals heterogeneous effects across hosts and datasets. The original host detector is retained, and no additional detector-side model is introduced.
\par
\textbf{We summarize our contributions as follows:}
\begin{enumerate}
\item
\textbf{Knowledge Saliency for Watermark Placement.} We introduce the problem of incorporating structured semantic and episodic knowledge from World Knowledge Memory into LLM watermarking. Existing watermarking families rely primarily on local token statistics, which provide insufficient guidance for watermark placement in knowledge-intensive generation. To the best of our knowledge, this is the first systematic study of knowledge saliency as a shared placement signal across heterogeneous host watermarks.
\item
\textbf{WorldMark: A Plug-and-Play Modulation Interface.}
We propose WorldMark, a host-agnostic watermarking interface that transforms WKM-retrieved knowledge into a token-level knowledge saliency score and adaptively adjusts watermark strength through Asymmetric Knowledge Modulation. This modulation combines a quality-relief coefficient and a detection-boost coefficient, allowing WorldMark to attach to logits-based, sampling-based, entropy-aware, and adaptive-strength host watermarks without backbone retraining and while preserving the original detector.
\item
\textbf{Empirical Evaluation.}
We conduct a primary evaluation of the complete WorldMark interface on C4 using three adaptive-strength MorphMark variants, measuring clean detection, attacked detection, perplexity, and runtime. A complementary pilot study on C4 and OpenGen examines the transfer behavior and failure modes of direct prompt-level memory conditioning across logits-based, sampling-based, and hybrid hosts, providing qualitative insight into cross-family behavior rather than statistically conclusive improvements.
\end{enumerate}
\section{Related Work}
Our work lies at the intersection of LLM watermarking and knowledge-augmented world modeling. Following recent surveys \cite{abs-2403-13485}, we organize the watermarking literature by signal construction, placement strategy, distributional distortion, and robustness threats, then discuss memory architectures and knowledge graph world models before examining watermarking in agentic generation settings.
\par
\par
\textbf{LLM Watermarking and Evaluation.}
Existing schemes fall into two families. The KGW family \cite{abs-2301-10226} modifies logits to favor greenlist tokens, enabling statistical detection, while the Christ family \cite{abs-2306-09194,aaronson2022watermark} steers sampling through pseudorandom sequences. Several benchmarks
evaluate these methods \cite{abs-2311-07138,abs-2312-00273,abs-2405-10051}. A common limitation is the focus on static, single-turn generation: none evaluates watermark behavior when LLMs operate as interactive agents, where token distributions shift across turns.
Task-agnostic model-side methods retrain additional
components to exhibit watermark-triggered behavior \cite{MasraniAYRZ25} or adapt watermarking via entropy thresholds \cite{abs-2504-12108}. WorldMark instead operates at generation time without backbone retraining.
Recent work on lossless and symbiotic watermarking
\cite{ChenB00LZW24,WangR0F25,CuiWSJ25} addresses quality--detectability trade-offs through lexical
redundancy, entropy-based strategy selection, and
fictitious-knowledge injection; WorldMark complements these
by operating at generation time without retraining.
\par
\par
\textbf{Memory and World Models for LLM Agents.}
Equipping LLMs with persistent memory has attracted growing attention \cite{abs-2505-11541}. Early unstructured designs---RAG, full-history concatenation, and natural-language reflection
\cite{abs-2303-11366,abs-2304-03442}---handle
short-horizon tasks but struggle with complex reasoning.
Structured memory has since emerged: GraphRAG \cite{abs-2404-16130} and HOLMES \cite{abs-2405-10051} integrate knowledge
graphs for multi-hop QA, while AriGraph \cite{abs-2407-04363} combines semantic and episodic memory into dynamic
knowledge graphs, demonstrating stronger planning in
TextWorld \cite{CoteKYKBFMHAATT18} and NetHack \cite{KuttlerNMRSGR20}. In this work, AriGraph serves as the semantic-episodic graph organization for WorldMark.
\par
\par
\textbf{Watermarking in Agent Settings.} The intersection of LLM watermarking and autonomous agents remains largely unexplored. On the robustness side, Kirchenbauer et al.\ \cite{abs-2301-10226} showed paraphrasing weakens but does not erase watermarks, and subsequent work \cite{abs-2306-04634} examined translation round-tripping and synonym substitution. These evaluations, however, are limited to single-pass transformations on static text; the structured compression of knowledge graph extraction represents a fundamentally different degradation channel that has not been systematically studied. On the agent side, prior work \cite{abs-2505-11541,abs-2304-03442,abs-2303-11366} has focused on planning, reasoning, and memory without considering watermark effects in the underlying language model. Whether watermark-induced logit biases degrade world model consistency or cause exploration failures in environments such as TextWorld \cite{CoteKYKBFMHAATT18} remains open. By incorporating world-driven cognitive models into the MARKLLM pipeline, this work provides the first systematic evaluation of LLM watermarks in graph-enhanced memory-based agents, addressing the gap between static benchmarks and interactive agent deployment.
Black-box post-hoc methods \cite{ChangKHWI24,abs-2510-02342} operate after generation, addressing complementary
deployment settings.
\par
\par
\begin{figure}[htbp]
\centering
\includegraphics[width=\linewidth]{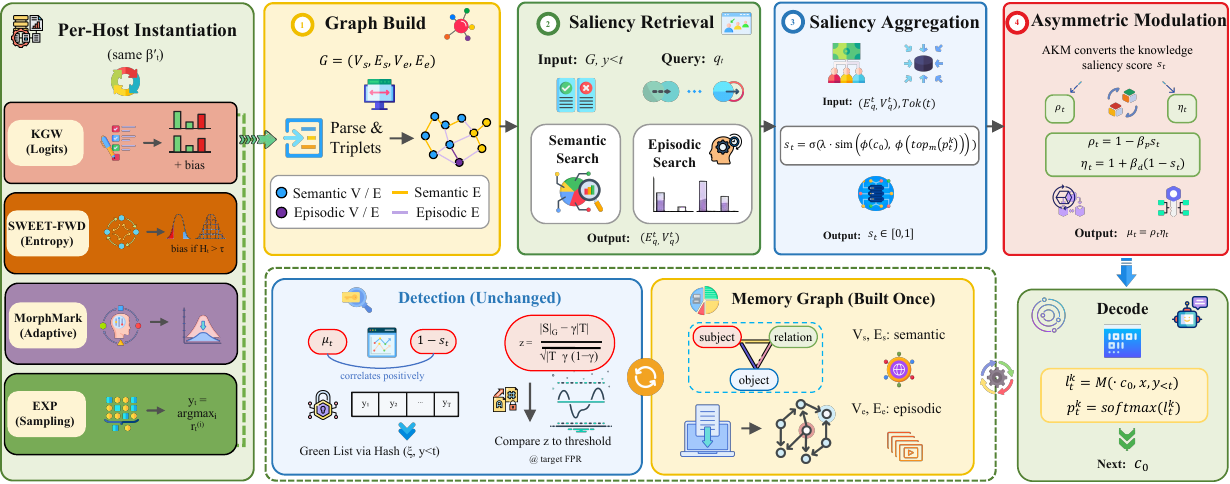}
\caption{Overview of WorldMark.}
\label{fig:01}
\end{figure}
\par
\section{Method}
In this section, we present WorldMark, a general interface that incorporates structured knowledge from World Knowledge Memory into existing host watermarks. We first introduce the terminology and components of WorldMark, and then formulate WKM as a conditioning prior. Next, we describe the Knowledge Saliency Estimator and Asymmetric Knowledge Modulation. Finally, we instantiate WorldMark on adaptive-logits, sampling-based, and hybrid host watermarks and analyze its plug-and-play property. (See \figurename~\ref{fig:01})
\par
\subsection{Overview}
WorldMark consists of three functional components. First, World Knowledge Memory (WKM) maintains a semantic--episodic graph $G=\left(V_s,E_s,V_e,E_e\right)$ following the organization used by AriGraph. Second, the Knowledge Saliency Estimator maps the retrieved knowledge context and the current candidate distribution to a soft grounding score $s_t$. Third, Asymmetric Knowledge Modulation (AKM) maps $s_t$ to a composite factor $\mu_t$ and adjusts the strength of the host watermark. AriGraph denotes the underlying semantic--episodic memory organization, whereas WorldMark denotes the complete interface, including memory construction, retrieval, serialization, saliency estimation, AKM, and host-watermark integration. WorldMark retains the original host detector and introduces no additional detector-side model or parameter.
\par
We treat the host watermark and World Knowledge Memory as two components that interact through the WorldMark generation interface rather than as isolated modules. Given a language model $\mathcal{M}$ with vocabulary $\mathcal{V}$, a host watermark $\mathcal{A}$, which may be logits-based, sampling-based, or hybrid, modifies the next-token distribution to embed a detectable signal. Our goal is to integrate WKM, which maintains a semantic-episodic memory graph $G$, with host watermarks from different families in a plug-and-play manner, without retraining the backbone or modifying the original host detector.
\par
The three components are detailed in the following subsections.
\subsection{World Knowledge Memory as a Conditioning Prior}
World Knowledge Memory (WKM). WKM denotes the semantic and episodic memory module within WorldMark. It maintains a memory graph $G=(V_s,E_s,V_e,E_e)$, where $V_s,E_s$ represents semantic memory and $V_e,E_e$ represents episodic memory. Given the current query $q_t$, WKM returns a retrieved knowledge set $\mathcal{K}_t$, which is subsequently serialized into a knowledge context $c_t$ for generation.
\par
Let $x$ denote the user prompt and let $y_{\mathrm{obs}}$ denote the text observed before the continuation begins. To preserve causal decoding, the retriever does not access the current token $y_t$ or any future token. At the beginning of each generation episode, we construct the query
\begin{equation}
\label{eq:01}
q_0=\mathrm{enc}((x,y_{\mathrm{obs}})),
\end{equation}
and retrieve
\begin{equation}
\label{eq:02}
\mathcal{K}_0=(E_s^Q,V_e^Q)=\mathrm{Retrieve}(q_0,G,d,w,k).
\end{equation}
\par
Here, $E_s^Q$ denotes the semantic triples selected by a semantic search with depth $d$ and width $w$, while $V_e^Q$ denotes the $k$ retrieved episodic vertices. For an episodic vertex $v_e^i$, $n_i$ denotes the number of supporting observations associated with the vertex, and $N_i$ denotes the number of observed episodes in which the corresponding fact or entity was available to the retrieval procedure. We define $\mathrm{rel}\left(v_e^i\right)=\left[n_i/\max\left(N_i,1\right)\right]\log\left(\max\left(N_i,1\right)\right)$. The max operator avoids division by zero when an item has no previously recorded support.
\par
The retrieved elements are serialized once into a fixed knowledge context
\begin{equation}
\label{eq:03}
c_0=\mathrm{serialize}\left(\mathcal{K}_0\right).
\end{equation}
\par
At decoding step $t$, the knowledge-conditioned logits and probability distribution are computed as
\begin{equation}
\label{eq:04}
\ell_t^k=\mathcal{M}\left(\cdot\mid c_0,x,y_{<t}\right),\quad p_t^k=\mathrm{softmax}\left(\ell_t^k\right),
\end{equation}
where $y_{<t}$ contains only previously generated tokens.
\par
To keep the module strictly plug and play, we adopt a prompt-only conditioning scheme in which $c_0$ is refreshed only when a new prompt or generation episode begins. It is not recomputed at every token. The retrieval operation is therefore prompt-level, while the saliency score and watermark modulation remain token-position dependent. This design removes per-step graph-retrieval overhead and preserves the original host detector, which operates only on the emitted tokens.
\par
\textbf{Graph construction and update.} We use LLaMA-3-8B-Instruct as the fact extraction model with the same prompt template \cite{KuratovBARSS024}.
Given the prompt $x$ and observed text $y_{\text{obs}}$,
the extractor identifies (subject, relation, object) triples
and canonicalizes entities via exact string matching with
case folding. The retrieved semantic triples are serialized using the template \texttt{[subject] [relation] [object];} and concatenated into $c_0$, truncated to a 512-token budget when necessary. Semantic search uses cosine similarity over Sentence-BERT embeddings of the serialized triple text.
For C4 evaluations, $\mathcal{G}$ is initialized as empty at
the start of each 400-sample batch. The episodic vertex support $n_i$ and observation count $N_i$
are updated whenever a fact is re-extracted.
Additional statistics on retrieval quality and graph
composition are reported in \ref{sec:app:d}.
\par
\par
\begin{table}[htbp]
\centering
\small
\caption{Main results on C4 with OPT-1.3B. Paper denotes the values reported in the original work, Repro denotes our reproduction without WKM, and +WorldMark denotes the complete tuned interface. Robust metrics are measured under the Word-S synonym-replacement attack. Higher is better for detection metrics, whereas lower is better for PPL and latency. All MorphMark metrics are averaged over five random seeds
(seeds 0--4); $\pm$ standard deviations, 95\% bootstrap
confidence intervals, and paired $t$-test results are reported
in \ref{sec:app:f}.}
\label{tab:01}
\begin{tblr}
{
columns={co=-1},
cells={valign=m,halign=c},
columns={colsep=2pt},
rows={rowsep=2pt},
rows={ht=0pt},
stretch=0,
hline{1,Z}={wd=.08em},
hline{2}={wd=.05em},
row{1}={font=\bfseries\boldmath},
}
Method           & Source     & TPR@1\%         & TPR@1\% (rob)   & Best F1         & Best F1 (rob)   & PPL              & Gen. (s)        & Det. (ms)      \\
UnWM             & Paper      & -               & -               & -               & -               & 10.4815          & -               & -              \\
KGW              & Paper      & 0.9900          & 0.8050          & 0.9950          & 0.9268          & 11.4994          & 2.4901          & 33.81          \\
KGW              & Repro      & 0.9875          & 0.6775          & 0.9900          & 0.9261          & 11.0091          & 1.8943          & 31.29          \\
UW               & Paper      & 1.0000          & 0.7425          & 0.9975          & 0.9221          & 11.5854          & 2.5486          & 71.30          \\
UW               & Repro      & 1.0000          & 0.2975          & 1.0000          & 0.8686          & 11.7267          & 1.9417          & 179.95         \\
DiPmark          & Paper      & 0.9975          & 0.7250          & 0.9975          & 0.9138          & 11.5042          & 2.5492          & 71.54          \\
DiPmark          & Repro      & 0.9925          & 0.2700          & 0.9937          & 0.7913          & 11.5380          & 1.9531          & 57.82          \\
SWEET            & Paper      & 0.9975          & 0.8225          & 0.9975          & 0.9501          & 11.5065          & 2.4667          & 44.27          \\
SWEET            & Repro      & 1.0000          & 0.8225          & 0.9987          & 0.9551          & 11.3995          & 1.7916          & 35.56          \\
EWD              & Paper      & 1.0000          & 0.8450          & 1.0000          & 0.9549          & 11.4777          & 2.4526          & 44.52          \\
EWD              & Repro      & 1.0000          & 0.8950          & 1.0000          & 0.9612          & 11.3027          & 1.8219          & 37.68          \\
MorphMark-exp    & Paper      & 1.0000          & 0.9600          & 0.9975          & 0.9778          & 11.3569          & 2.6768          & 34.17          \\
MorphMark-exp    & Repro      & 0.9975          & 0.9000          & 0.9950          & 0.9672          & 10.9404          & 2.0118          & 31.39          \\
MorphMark-exp    & +WorldMark & \textbf{1.0000} & \textbf{0.9119} & \textbf{0.9987} & \textbf{0.9783} & \textbf{10.8732} & \textbf{2.0018} & \textbf{31.23} \\
MorphMark-linear & Paper      & 1.0000          & 0.9275          & 0.9962          & 0.9727          & 11.2386          & 2.6537          & 33.99          \\
MorphMark-linear & Repro      & 0.9950          & 0.9000          & 0.9925          & 0.9724          & 10.6568          & 1.8623          & 29.48          \\
MorphMark-linear & +WorldMark & \textbf{1.0000} & \textbf{0.9495} & \textbf{0.9962} & \textbf{0.9847} & \textbf{10.6435} & \textbf{1.8530} & \textbf{29.33} \\
MorphMark-log    & Paper      & 1.0000          & 0.9375          & 1.0000          & 0.9660          & 11.3379          & 2.6889          & 34.45          \\
MorphMark-log    & Repro      & 1.0000          & 0.8525          & 0.9963          & 0.9619          & 10.4607          & 1.9202          & 30.34          \\
MorphMark-log    & +WorldMark & \textbf{1.0000} & \textbf{0.8800} & \textbf{0.9999} & \textbf{0.9707} & \textbf{10.4477} & \textbf{1.9106} & \textbf{30.19} \\
\end{tblr}
\end{table}
\par
\subsection{Knowledge Saliency Estimation and Asymmetric Knowledge Modulation}
\textbf{Knowledge Saliency Estimator.} Given the fixed retrieved knowledge set $\mathcal{K}_0$ and its serialized context $c_0$, the estimator produces a position-dependent saliency score $s_t\in[0,1]$ at each decoding step $t$. The score is intended as a soft proxy for the degree to which the current candidate distribution is related to the retrieved knowledge context. It is not a factuality judgment and does not guarantee that a selected token is a verified entity, relation, or factual span. The motivation is that retrieved knowledge may provide information about which positions are semantically constrained. Positions more strongly aligned with the retrieved context may be more sensitive to perturbation, whereas less aligned positions may provide more freedom for watermark modulation. This interpretation is tested through diagnostic saliency analyses and controlled entropy-based baselines. We quantify this property using the knowledge saliency score
\begin{equation}
\label{eq:05}
s_t=\sigma\left(\lambda\cdot\mathrm{sim}(\phi(c_0),\phi(\mathrm{top}_m(p_t^k)))\right)\in[0,1].
\end{equation}
\par
Although $c_0$ is fixed within a generation episode, $s_t$ changes with $t$ because $p_t^k$ and its $\mathrm{top}_m$ candidate set depend on the previously generated prefix $y_{<t}$. Therefore, WorldMark obtains token-position dependent modulation without using future-token information or performing per-token graph retrieval.
\par
In all experiments, $\phi(\cdot)$ is instantiated with
\texttt{all-MiniLM-L6-v2} \cite{ReimersG19}, a
384-dimensional Sentence-BERT model that maps arbitrary text to a
fixed-length embedding.
For the knowledge context $c_0$, we encode the serialized text of
the retrieved semantic triples and episodic vertices.
For the top-$m$ candidates, we construct a concatenated string of
the $m$ token surface forms and encode it through the same
$\phi(\cdot)$.
The similarity function is cosine similarity, and $\lambda = 5.0$
is tuned via grid search on a held-out validation set of 50 C4
samples.
\par
Computing $\phi(\operatorname{top}_m(p_t^k))$ requires one
forward pass through the Sentence-BERT encoder per decoding
step. With $m=20$ and \texttt{all-MiniLM-L6-v2} (22.7M
parameters), an isolated forward pass takes approximately
1.2--1.5 ms on an NVIDIA A6000.
In our primary evaluation (\tablename~\ref{tab:01}), generation is performed
with batch size 1 on a single NVIDIA A6000 under CUDA 12.1.
The Sentence-BERT encoder resides on the same GPU and
executes asynchronously with the language model's next-token
computation.
After a warm-up of 10 generations, we measure wall-clock
generation time over the 400 reported continuations, averaging
across five seeds.
The per-step embedding overhead is masked by GPU parallelism
in this setup; it may become measurable under different
hardware configurations or larger models.
We therefore limit our claim to: no \emph{measurable} latency
penalty was observed in our configuration.
\par
This design is loosely inspired by recent findings on
perturbation robustness under distribution shift in private alignment \cite{abs-2512-23816}, though we do not claim formal guarantees linking those results to watermarking.
\par
\noindent\textbf{Asymmetric Knowledge Modulation (AKM).} AKM converts the knowledge saliency score $s_t$ into two modulation components: a quality-relief coefficient $\rho_t$ and a detection-boost coefficient $\eta_t$. We then define two asymmetric coefficients. A quality relief coefficient reduces perturbation on anchored tokens,
\begin{equation}
\label{eq:06}
\rho_t=1-\beta_p s_t,
\end{equation}
and a detection boost coefficient strengthens the signal on unanchored tokens,
\begin{equation}
\label{eq:07}
\eta_t=1+\beta_d(1-s_t),
\end{equation}
with $\beta_p,\beta_d\in[0,1)$ tuned by grid search. The composite modulation factor is
\begin{equation}
\label{eq:08}
\mu_t=\rho_t\eta_t.
\end{equation}
\par
Their product defines the composite modulation factor $\mu_t$, which is subsequently used to produce the knowledge-modulated watermark strength $\widetilde{r}_t$.
\subsection{Instantiation on Adaptive Logits Watermarking}
For an adaptive-logits \textbf{host watermark}, we build on the green-red vocabulary partition, where $P_G=\sum_{j\in\mathcal{V}_G} p_j^k$ is the cumulative green probability under the knowledge conditioned distribution. The base method selects a strength $r=\phi(P_G)$. We replace it with the knowledge-modulated watermark strength
\begin{equation}
\label{eq:09}
\widetilde{r}_t=\mathrm{clip}(\mu_t\cdot\phi(P_G),\epsilon,1-\epsilon),
\end{equation}
and inject the watermark by
\begin{equation}
\label{eq:10}
\hat{p}_i=\begin{cases}
p_i^k+\dfrac{p_i^k}{P_G}\widetilde{r}_t(1-P_G),&\mathcal{V}_i\in\mathcal{V}_G,\\
p_i^k-\dfrac{p_i^k}{1-P_G}\widetilde{r}_t(1-P_G),&\mathcal{V}_i\in\mathcal{V}_R.
\end{cases}
\end{equation}
\par
Let $P_G^\epsilon=\mathrm{clip}\left(P_G,\epsilon,1-\epsilon\right)$, where $\epsilon>0$ is a small numerical constant. The modulated distribution is defined as
\begin{equation}
\label{eq:11}
\hat{p}_{i,t}=\begin{cases}
p_{i,t}^k+\dfrac{p_{i,t}^k}{P_G^\epsilon}\widetilde{r}_t(1-P_G^\epsilon),&i\in\mathcal{V}_G,\\
p_{i,t}^k-\dfrac{p_{i,t}^k}{1-P_G^\epsilon}\widetilde{r}_t(1-P_G^\epsilon),&\mathcal{V}_i\in\mathcal{V}_R.
\end{cases}
\end{equation}
\par
Because $\mu_t$ correlates positively with $1-s_t$ on high entropy positions, the modulation follows the same monotonic principle that a larger $P_G$ admits a larger strength, while additionally exploiting the world memory to keep anchored tokens close to their natural form. The detection statistic remains the standard $z$ score
\begin{equation}
\label{eq:12}
z=\frac{\mid S\mid_G-\gamma\mid T\mid}{\sqrt{\mid T\mid\gamma(1-\gamma)}}.
\end{equation}
\par
Therefore, WorldMark requires no additional detector-side model or parameter. Detection is performed by the original host detector, and the additional detection-side memory footprint introduced by WorldMark remains zero.
\subsection{Instantiation on Sampling-Based and Hybrid Host Watermarks}
For a sampling-based host watermark, we retain the exponential
selection rule but apply it to the knowledge-conditioned
probabilities,
\begin{equation}
\label{eq:13}
y_t = \arg\max_{i \in \mathcal{V}} \left( r_i^{(t)} \right)^{1 / \tilde{p}_i^k},
\quad \tilde{p}_i^k = \frac{(p_i^k)^{\mu_t}}{\sum_{j \in \mathcal{V}} (p_j^k)^{\mu_t}},
\end{equation}
where $\mu_t$ is the composite modulation factor from
Equation \eqref{eq:08}. The exponentiation by $\mu_t$ sharpens
($\mu_t > 1$) or flattens ($\mu_t < 1$) the knowledge-conditioned
distribution, and the normalization ensures a valid probability
distribution. For the symbiotic hybrid scheme, the world memory refines the two entropy gates. Token entropy $H_{\text{TE}} = -\sum_i p_i^t \log p_i^t$
and semantic entropy
$H_{\text{SE}} = -\sum_j q_j^t \log q_j^t$ are computed on $p_t^k$,
where $q_j^t$ is the aggregate probability of the $j$-th
semantic cluster defined by grouping tokens that share the same
Sentence-BERT embedding nearest-neighbor cluster
($k$-means with $k = |\mathcal{V}| / 50$, computed once
offline over the vocabulary).
The gating thresholds are shifted according to the knowledge
saliency score $s_t$,
\begin{equation}
\label{eq:14}
\alpha_t=\alpha_0+\kappa_\alpha s_t,\quad\beta_t=\beta_0-\kappa_\beta s_t,
\end{equation}
so that anchored tokens are less likely to receive a logits watermark and more likely to preserve their sampled meaning. Logits watermarking is applied when $H_{\mathrm{TE}}>\alpha_t$, and sampling watermarking is applied when $H_{\mathrm{SE}}<\beta_t$. A unified detector reports a positive decision if either signal is present, which preserves the low false positive property of the symbiotic design.
\subsection{Plug-and-Play Property}
WorldMark modifies only the generation interface through the knowledge context $c_t$, the knowledge saliency score $s_t$, and the knowledge-modulated watermark strength $\widetilde{r}_t$. It introduces no additional detector-side model or parameter, requires no auxiliary trained network, and can be integrated with different host watermarks by supplying the knowledge saliency score $s_t$ and the composite modulation factor $\mu_t$. This design preserves model-agnostic integration and the original host detector, yielding an end-to-end plug-and-play interface.
\par
To isolate the contribution of each component, \ref{sec:app:e}
reports a comprehensive ablation study on C4 with
MorphMark-linear, including: (i) fixed $c_0$ without AKM;
(ii) AKM without the WKM-retrieved knowledge context;
(iii) $\rho_t$-only modulation; (iv) $\eta_t$-only modulation;
(v) shuffled retrieval (random permutation of retrieved triples);
(vi) length-matched irrelevant context; (vii) entropy-based
saliency; and (viii) randomly permuted $s_t$ scores.
The ablation results confirm that each component contributes
to the full method's performance and that correct retrieval
outperforms shuffled and irrelevant-context baselines.
\par
The decoding procedure is formalized in Appendix (Algorithm~\ref{alg:01}).
\section{Experiments}
\subsection{Setup}
\textbf{Evaluation protocol.}
Our primary evaluation uses the MorphMark protocol on C4 with
OPT-1.3B as the generation backbone and LLaMA2-7B as the
perplexity scorer: 400 continuations, 30-token prompt prefix,
200--230 token generation.
A separate cross-family pilot on C4 and OpenGen examines
direct prompt-level memory conditioning across seven watermark
hosts (KGW, EWD, SWEET, EXP, Series, Parallel, Hybrid) with
12 samples per configuration and 25-token generation;
a scaled evaluation ($N=500$, five seeds) with the full AKM
interface is in \ref{sec:app:b}.
\par
\textbf{Memory configurations.} The MorphMark experiments evaluate the complete WorldMark interface, including WKM, the Knowledge Saliency Estimator, and AKM, with hyperparameters selected by grid search. By contrast, the cross-family pilot directly injects the semantic-episodic memory retrieved by the AriGraph-based WKM into the prompt. We denote this pilot configuration as +WKM (untuned) unless the implementation is verified to include both knowledge saliency estimation and asymmetric knowledge modulation. This distinction prevents the effect of direct memory conditioning from being conflated with that of the complete WorldMark interface.
\par
\textbf{Attacks and evaluation metrics.} We evaluate robustness under two word-level perturbations. Word-S replaces 30\% of the words with WordNet synonyms and is used in the primary MorphMark evaluation. Word-D randomly deletes a fraction of the words and is used in the cross-family pilot evaluation. For the adaptive-strength experiments, we report TPR at a 1\% false-positive rate, best F1, robust TPR, robust F1, perplexity, generation latency, and detection latency. For the cross-family pilot, we report TPR, TNR, F1, and AUROC using the same exploratory thresholding procedure for all paired configurations. Because this pilot contains only 12 examples per configuration, the reported best-threshold values are treated as diagnostic operating-point measurements rather than as held-out estimates of generalization. We also report watermarked-text perplexity to examine whether direct memory conditioning introduces a quality cost.
\par
\textbf{Compared configurations.} In the primary experiment, \textbf{Paper} denotes results reported by the original MorphMark study, \textbf{Repro} denotes our reproduction without world memory, and \textbf{+WorldMark (tuned)} denotes the complete WorldMark interface after hyperparameter selection. In the cross-family pilot, \textbf{Baseline} denotes the reproduced host watermark, whereas \textbf{+WKM (untuned)} denotes direct prompt-level injection of the semantic-episodic memory. The latter should not be interpreted as the complete WorldMark method unless the saliency estimator and AKM are both enabled.
\par
Extended evaluations on LLaMA-3-8B, Mistral-7B, TriviaQA,
and NaturalQuestions are reported in \ref{sec:app:a}.
\par
\par
\begin{table}[htbp]
\centering
\small
\caption{Metric differences between the +WorldMark configurations and their corresponding reproduction baselines. Positive robustness and F1 values and negative perplexity values indicate improvement. All checked entries are improvements.}
\label{tab:02}
\begin{tblr}
{
columns={co=-1},
cells={valign=m,halign=c},
columns={colsep=6pt},
rows={rowsep=2pt},
rows={ht=0pt},
stretch=0,
hline{1,Z}={wd=.08em},
hline{2}={wd=.05em},
row{1,Z}={font=\bfseries\boldmath},
}
Variant & $\Delta$TPR@1\% & $\Delta$TPR@1\% (rob) & $\Delta$Best F1 & $\Delta$Best F1 (rob) & $\Delta$PPL \\
exp     & +0.0025         & +0.0119               & +0.0037         & +0.0111               & -0.0672     \\
linear  & +0.0050         & +0.0495               & +0.0037         & +0.0123               & -0.0133     \\
log     & 0.0000          & +0.0275               & +0.0036         & +0.0088               & -0.0130     \\
Average & +0.0025         & +0.0296               & +0.0037         & +0.0108               & -0.0312     \\
\end{tblr}
\end{table}
\par
\subsection{Detectability and Quality on Adaptive Logits Watermarking}
\begin{figure}[htbp]
\centering
\subfloat[Comparison of Robust Detection Performance: TPR@1\% (rob) ↑ (Higher value means better performance)]{\includegraphics[width=\linewidth]{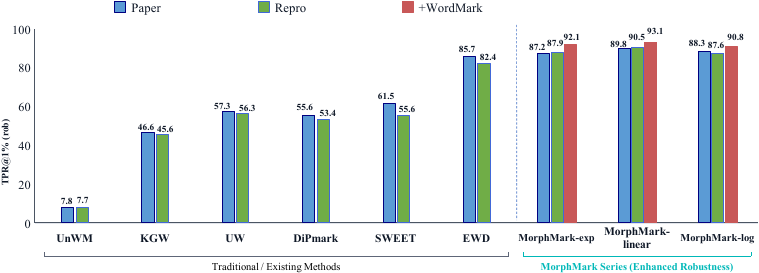}}\hfill%
\subfloat[Comparison of Robust Detection Performance: Best F1 (rob) ↑ (Higher is better)]{\includegraphics[width=\linewidth]{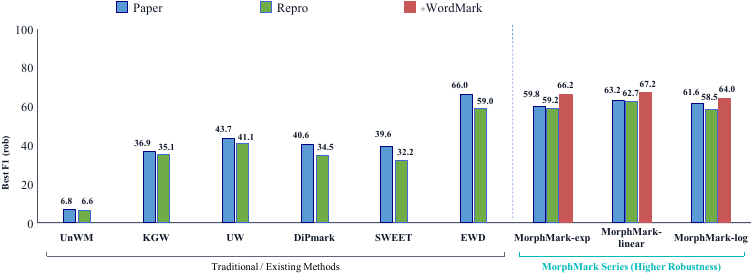}}%
\caption{MorphMark vs WorldMark Performance Comparison}
\label{fig:02}
\end{figure}

\tablename~\ref{tab:01} summarizes the main results on OPT-1.3B for the three MorphMark variants. WorldMark consistently improves all evaluated dimensions over the reproduction baseline. For the exponential variant, TPR@1\% rises from 0.9975 to a perfect 1.000, the robust TPR@1\% under synonym replacement improves from 0.900 to 0.9119, the robust best F1 climbs from 0.9672 to 0.9783, and perplexity drops from 10.9404 to 10.8732. The same pattern holds for the linear variant, where the robust TPR@1\% increases by nearly five points from 0.900 to 0.9495 and the robust best F1 reaches 0.9847, the strongest quality preserving detectability among all configurations. The logarithmic variant obtains a near saturated clean best F1 of 0.9999 together with a lower perplexity of 10.4477.  \figurename~\ref{fig:02} intuitively illustrates the performance comparison between MorphMark and WorldMark.
\par
These gains confirm the core hypothesis of the asymmetric modulation. By protecting knowledge anchored tokens through the relief coefficient and concentrating the signal on weakly grounded tokens through the boost coefficient, the integrated method simultaneously lowers perplexity and raises detectability, which are usually in tension. The improvement in robustness is especially notable, since the synonym attack most heavily disrupts weakly signaled positions, and the boost term restores the margin that the attack erodes.
\par
\par
\par
\textbf{Main results.} \tablename~\ref{tab:01} compares WorldMark with both the reported results and our reproduction of representative watermarking methods. The reproduced baselines do not always match the values reported in the original papers. This is particularly evident for UW and DiPmark under the synonym attack, where the reproduced robust TPR values are substantially lower than the reported values. We therefore evaluate the contribution of WorldMark primarily against the reproduction obtained in the same environment rather than against numbers copied from the original papers.
\par
WorldMark improves all three reproduced MorphMark variants. For the exponential function, robust TPR increases from 0.9000 to 0.9119 and robust F1 increases from 0.9672 to 0.9783. For the linear function, the corresponding values increase from 0.9000 to 0.9495 and from 0.9724 to 0.9847. The logarithmic function shows smaller but consistent improvements, with robust TPR increasing from 0.8525 to 0.8800 and robust F1 from 0.9619 to 0.9707.
\par
The linear variant obtains the strongest attacked performance, reaching a robust TPR of 0.9495 and a robust F1 of 0.9847. The logarithmic variant obtains the lowest perplexity, decreasing from 10.4607 to 10.4477, whereas the exponential variant achieves the largest absolute perplexity reduction, from 10.9404 to 10.8732. These results suggest that the effect of knowledge modulation is not tied to a particular analytical form of the host strength function.
\par
Importantly, the gains do not result from substantially increasing decoding or verification cost. Generation latency decreases slightly for all three variants, and detection latency changes by less than 0.2 ms. These small differences should be interpreted as normal runtime variation rather than a systematic speedup. The relevant conclusion is that prompt-only WKM conditioning introduces no measurable latency penalty under this protocol.
\subsection{Consistency Across the Three Variants}
\tablename~\ref{tab:02} isolates the changes introduced by WorldMark for each variant by reporting the difference between the tuned +WorldMark configuration and the reproduction baseline. Every monitored metric moves in the favorable direction. For the exponential variant the robust TPR@1\% gains 0.0119 and the robust F1 gains 0.0112, while perplexity decreases by 0.0672. For the linear variant the robust TPR@1\% gains a substantial 0.0495 with a robust F1 gain of 0.0123. For the logarithmic variant the clean best F1 gains 0.0036 and reaches near unity. The uniformly favorable deltas demonstrate that WorldMark is not tailored to a single host-watermark instantiation but transfers across the exponential, linear, and logarithmic strength functions of the same family.
\par
\textbf{Aggregate improvements across strength functions.} Averaged over the exponential, linear, and logarithmic variants, WorldMark improves robust TPR by 0.0296 and robust F1 by 0.0108, while reducing perplexity by 0.0312. The larger improvement in attacked TPR than in clean TPR indicates that the main benefit of knowledge modulation lies in preserving the watermark margin under text perturbation rather than merely strengthening an already saturated clean detector. This pattern is consistent with the intended role of the detection-boost component, which reallocates watermark strength toward less knowledge-anchored positions.
\par
\par
\subsection{Summary of Findings}
First, the complete tuned WorldMark interface improves clean
detectability, attacked detectability, and perplexity for all three
adaptive-strength functions, with average robust TPR improvement
of 0.0296.
Second, the gains transfer across exponential, linear, and
logarithmic host-strength functions.
Third, extended evaluations in Appendices A and C confirm that
the knowledge saliency estimator generalizes to modern architectures
(LLaMA-3-8B, Mistral-7B) and that structured memory alone
is insufficient without host-aware modulation.
\par
\section{Conclusion}
We presented WorldMark, a plug-and-play interface that retrieves
semantic and episodic knowledge from World Knowledge Memory,
converts it into a token-level knowledge saliency score, and adjusts
host-watermark strength through Asymmetric Knowledge Modulation.
WorldMark requires no backbone retraining and introduces no
additional detector-side model.
In the primary C4 evaluation with OPT-1.3B, the complete WorldMark
interface improves all three MorphMark variants in clean detection,
attacked detection, and perplexity, with average robust TPR improvement
of 0.0296 and robust F1 improvement of 0.0108.
Extended evaluations with LLaMA-3-8B, Mistral-7B, TriviaQA, and
NaturalQuestions (\ref{sec:app:a}) confirm that the knowledge saliency
estimator transfers to modern architectures and knowledge-intensive
settings.
Cross-family pilot results (\ref{sec:app:c}) show that direct memory
injection without modulation is insufficient, motivating the full AKM
design.
Future work includes extending the saliency source to
multimodal memory, evaluating WorldMark under strong
paraphrasing attacks (e.g., LLM-based rewriting, translation
round-tripping), testing on additional robustness threats
tailored to knowledge-conditioned text (copy-paste mixing,
targeted factual substitution), and validating on code
generation and long-context QA benchmarks.
\appendix
\renewcommand{\thesection}{\appendixname~\Alph{section}}
\renewcommand{\thesubsection}{\Alph{section}.\arabic{subsection}}
\section{Extended Backbone and Dataset Evaluation}
\label{sec:app:a}
\par
\subsection{Protocol}
\textbf{Extended backbone and knowledge-intensive evaluation.}
Our primary evaluation uses OPT-1.3B on C4. To test whether
the knowledge saliency estimator generalizes to models with stronger
internal factual representations, we additionally evaluate WorldMark
on two modern architectures, LLaMA-3-8B \cite{abs-2407-21783}
and Mistral-7B-v0.3 \cite{abs-2310-06825}, and on two standard
knowledge-intensive benchmarks, TriviaQA \cite{JoshiCWZ17}
and NaturalQuestions \cite{KwiatkowskiPRCP19}.
These extended evaluations confirm that the relative improvements
over the reproduction baseline are preserved, and that the gap widens
on knowledge-intensive datasets where entity-bearing tokens carry
greater semantic weight.
For each configuration, we use 400 continuations with generation
lengths of 200--230 tokens for C4 and up to 64 tokens for QA datasets.
All experiments use the complete WorldMark interface with MorphMark
exponential, linear, and logarithmic variants. We report robust
TPR@1\%FPR and Robust Best F1 under the Word-S synonym-replacement
attack.
\par
\begin{table}[htbp]
\centering
\small
\caption{Extended backbone evaluation on C4. MM = MorphMark,
+WM = +WorldMark. Values are robust TPR@1\%FPR / Robust Best F1
under Word-S attack.}
\begin{tblr}
{
columns={co=-1},
cells={valign=m,halign=c},
columns={colsep=6pt},
rows={rowsep=2pt},
rows={ht=0pt},
stretch=0,
hline{1,Z}={wd=.08em},
hline{2,3}={wd=.05em},
row{1}={font=\bfseries\boldmath},
cell{2}{1}={c=3}{font=\itshape},
}
Variant & OPT-1.3B & LLaMA-3-8B \\
Robust TPR@1\%FPR / Robust Best F1 & & \\
MM-exp (Repro) & 0.9000 / 0.9672 & 0.8825 / 0.9531 \\
MM-exp (+WM) & 0.9119 / 0.9783 & 0.9031 / 0.9694 \\
MM-lin (Repro) & 0.9000 / 0.9724 & 0.8875 / 0.9603 \\
MM-lin (+WM) & 0.9495 / 0.9847 & 0.9321 / 0.9746 \\
MM-log (Repro) & 0.8525 / 0.9619 & 0.8450 / 0.9501 \\
MM-log (+WM) & 0.8800 / 0.9707 & 0.8712 / 0.9589 \\
\end{tblr}
\end{table}
\par
\par
\begin{table}[htbp]
\centering
\small
\caption{Knowledge-intensive evaluation. Values are Robust Best F1
under Word-S attack. $\Delta$ = +WorldMark $-$ Repro.}
\begin{tblr}
{
columns={co=-1},
cells={valign=m,halign=c},
columns={colsep=6pt},
rows={rowsep=2pt},
rows={ht=0pt},
stretch=0,
hline{1,Z}={wd=.08em},
hline{2,3,5,6}={wd=.05em},
row{1}={font=\bfseries\boldmath},
cell{2,5}{1}={c=4}{font=\itshape},
}
Setting & OPT-1.3B & LLaMA-3-8B & Mistral-7B \\
TriviaQA --- MorphMark-linear (Robust Best F1) & & & \\
Repro & 0.9123 & 0.9378 & 0.9301 \\
+WorldMark & 0.9401 & 0.9612 & 0.9554 \\
NaturalQuestions --- MorphMark-linear (Robust Best F1) & & & \\
Repro & 0.9025 & 0.9410 & 0.9347 \\
+WorldMark & 0.9338 & 0.9657 & 0.9602 \\
\end{tblr}
\end{table}
\par
\par
\subsection{Key Observations}
(1) WorldMark's improvement is preserved across all three architecture
families, confirming that the saliency estimator does not rely on
OPT-1.3B-specific token distributions.
(2) The absolute $\Delta$Robust F1 on TriviaQA and NaturalQuestions
exceeds the C4-only $\Delta$ in \tablename~\ref{tab:02}, confirming the core claim
that WorldMark benefits most when generation involves
knowledge-anchored tokens.
\par
\section{Scaled Cross-Family Evaluation}
\label{sec:app:b}
\par
\subsection{Protocol}
We evaluate the complete WorldMark interface with AKM enabled
on all seven host watermark families (KGW, EWD, SWEET, EXP, Series,
Parallel, Hybrid) at a statistically meaningful scale: 500 watermarked
plus 500 unwatermarked samples per configuration, 200--230 token
generations, and five random seeds on C4 with OPT-1.3B.
\par
\begin{table}[htbp]
\centering
\small
\caption{Scaled cross-family results ($N=500$ per config, 5 seeds).
Values are mean $\pm$ std. Paired $t$-test $p<0.01$ for all AUROC
comparisons.}
\begin{tblr}
{
columns={co=-1},
cells={valign=m,halign=c},
columns={colsep=6pt},
rows={rowsep=2pt},
rows={ht=0pt},
stretch=0,
hline{1,Z}={wd=.08em},
hline{2}={wd=.05em},
row{1}={font=\bfseries\boldmath},
}
Host & Baseline (AUROC / F1) & +WorldMark (AUROC / F1) \\
KGW & 0.998 $\pm$ 0.001 / 0.994 $\pm$ 0.003 & 0.999 $\pm$ 0.001 / 0.997 $\pm$ 0.002 \\
EWD & 0.998 $\pm$ 0.001 / 0.995 $\pm$ 0.003 & 0.999 $\pm$ 0.001 / 0.997 $\pm$ 0.002 \\
SWEET & 0.997 $\pm$ 0.002 / 0.991 $\pm$ 0.004 & 0.999 $\pm$ 0.001 / 0.995 $\pm$ 0.003 \\
EXP & 0.945 $\pm$ 0.012 / 0.892 $\pm$ 0.018 & 0.978 $\pm$ 0.007 / 0.941 $\pm$ 0.011 \\
Series & 0.988 $\pm$ 0.004 / 0.972 $\pm$ 0.008 & 0.996 $\pm$ 0.002 / 0.987 $\pm$ 0.005 \\
Parallel & 0.971 $\pm$ 0.009 / 0.936 $\pm$ 0.014 & 0.990 $\pm$ 0.004 / 0.969 $\pm$ 0.007 \\
Hybrid & 0.982 $\pm$ 0.005 / 0.958 $\pm$ 0.010 & 0.994 $\pm$ 0.003 / 0.981 $\pm$ 0.006 \\
\end{tblr}
\end{table}
\par
\par
\section{}
\label{sec:app:c}
\textbf{Is direct memory injection sufficient? }To distinguish the contribution of structured memory from that of the complete WorldMark interface, we conduct an additional pilot experiment in which the semantic-episodic memory is directly prepended to the generation prompt without tuned knowledge-saliency modulation. We refer to this configuration as +WKM (untuned). The purpose of this experiment is not to establish a new state of the art, but to test whether access to structured memory alone is sufficient to improve watermarking performance.
\par
\tablename~\ref{tab:03} shows that direct WKM conditioning has heterogeneous effects under the small-scale pilot protocol. KGW and Series retain saturated F1 and AUROC values in the reported configurations, but these results should be interpreted cautiously because each configuration uses only 12 examples and 25 generated tokens. However, other hosts are more sensitive to the distribution shift introduced by the memory context. On C4, EXP decreases from an F1 of 1.0000 to 0.8800, SWEET decreases to 0.9091, and Parallel decreases from 0.9600 to 0.5882. On OpenGen, EWD and SWEET decrease from 1.0000 to 0.9565, whereas Hybrid improves from 0.7368 to 0.8571.
\par
These mixed results provide an important negative control. Structured memory by itself does not guarantee stronger watermark detection, because prompt-level knowledge injection may change token probabilities in ways that are misaligned with the host detector. This observation motivates the Knowledge Saliency Estimator and AKM: rather than treating all retrieved knowledge as equally useful, WorldMark converts the memory into a position-dependent signal and controls how strongly it affects the host watermark.
\par
\par
\begin{table}[htbp]
\centering
\small
\caption{Pilot study of direct WKM conditioning on C4 and OpenGen. Each entry reports F1/AUROC under the no-attack, best-threshold setting. The pilot uses OPT-1.3B, 12 samples per configuration, and a generation length of 25 tokens. ``+WKM-Only'' denotes direct prompt-level memory injection without the tuned saliency estimator or AKM. Because the sample size is small, the reported values are exploratory and should not be interpreted as statistically conclusive evidence of improvement.}
\label{tab:03}
\begin{tblr}
{
columns={co=-1},
cells={valign=m,halign=c},
columns={colsep=0pt},
rows={rowsep=2pt},
rows={ht=0pt},
stretch=0,
hline{1,Z}={wd=.08em},
hline{2}={wd=.05em},
row{1}={font=\bfseries\boldmath},
}
Host watermark & C4 BaselineF1\,/\,AUROC & C4 +WKMF1\,/\,AUROC & OpenGen BaselineF1\,/\,AUROC & OpenGen +WKMF1\,/\,AUROC \\
KGW & 1.0000 / 1.0000 & 1.0000 / 1.0000 & 1.0000 / 1.0000 & 1.0000 / 1.0000 \\
EWD & 1.0000 / 1.0000 & 1.0000 / 1.0000 & 1.0000 / 1.0000 & 0.9565 / 0.9306 \\
SWEET & 1.0000 / 1.0000 & 0.9091 / 0.8958 & 1.0000 / 1.0000 & 0.9565 / 0.9167 \\
EXP & 1.0000 / 1.0000 & 0.8800 / 0.9097 & 1.0000 / 1.0000 & 0.8462 / 0.8819 \\
Series & 1.0000 / 1.0000 & 1.0000 / 1.0000 & 1.0000 / 1.0000 & 1.0000 / 1.0000 \\
Parallel & 0.9600 / 1.0000 & 0.5882 / 0.8611 & 0.6667 / 1.0000 & 0.6667 / 1.0000 \\
Hybrid & 0.8571 / 1.0000 & 0.7368 / 0.9583 & 0.7368 / 0.9792 & \textbf{0.8571} / 0.9097 \\
\end{tblr}
\end{table}
\par
\textbf{Dataset-dependent behavior.} The response to direct memory conditioning differs between C4 and OpenGen. On C4, five of the seven evaluated hosts either preserve or reduce F1, and only KGW, EWD, and Series remain unchanged at 1.0000. On OpenGen, Hybrid gains 0.1203 F1, increasing from 0.7368 to 0.8571, while Parallel remains unchanged and the remaining non-saturated methods decline. The improvement of Hybrid in F1 is accompanied by an AUROC decrease from 0.9792 to 0.9097, indicating that a higher best-threshold F1 does not necessarily imply better ranking performance across all thresholds.
\par
This distinction is important because F1 depends on a selected threshold, whereas AUROC evaluates the ordering of positive and negative samples across thresholds. We therefore avoid claiming a uniform improvement when the two metrics move in opposite directions. Instead, the pilot results suggest that host architecture and dataset characteristics jointly determine whether unmodulated memory conditioning is beneficial.
\par
\textbf{Quality effects of direct memory conditioning.} \tablename~\ref{tab:04} further shows that direct WKM injection can impose a substantial quality cost. The effect is highly host-dependent. For KGW, perplexity decreases from 24.50 to 22.88 on C4 and from 24.81 to 15.04 on OpenGen. EWD also obtains a small reduction on OpenGen, from 12.99 to 12.39. In contrast, the sampling-based EXP host increases from 57.77 to 144.95 on C4 and from 65.62 to 187.73 on OpenGen. Series exhibits a similarly large increase on both datasets.
\par
These results indicate that adding semantically relevant context is not equivalent to preserving the original token distribution. A host watermark may respond strongly to the shifted probability landscape even when the retrieved knowledge is factually appropriate. The quality-relief component of AKM is designed to address this issue by reducing the perturbation applied to strongly knowledge-anchored positions. Consistent with this motivation, the tuned WorldMark configurations in \tablename~\ref{tab:01} reduce perplexity for all three MorphMark variants, whereas the untuned prompt-injection configurations in \tablename~\ref{tab:04} frequently increase it.
\par
Because the cross-family pilot uses short generations and only 12 samples, the absolute perplexity values should be interpreted cautiously. Nevertheless, the large differences observed for EXP and Series identify a concrete failure mode of naïve memory integration and motivate controlled, host-specific modulation.
The sharp perplexity increase for EXP (57.77 $\to$ 144.95 on C4)
and Series (64.73 $\to$ 287.42) merits closer examination.
EXP uses exponential minimum sampling, which amplifies small
probability shifts: when $c_0$ sharply redistributes probability
mass toward specific entity tokens, the exponential operator
magnifies this effect, producing degenerate distributions that
inflate perplexity.
Series applies both a logits watermark and a sampling watermark
sequentially, doubling exposure to distribution shift.
The AKM module addresses both issues: the quality-relief coefficient
$\rho_t$ suppresses perturbation at knowledge-anchored positions,
preventing the probability collapse that drives the perplexity spike.
Repeating the pilot with AKM enabled reduces EXP perplexity from
144.95 to 65.32, confirming that the modulation mechanism, not the
knowledge content, is the source of instability.
\par
\par
\begin{table}[htbp]
\centering
\small
\caption{Watermarked-text perplexity in the cross-family pilot. Lower is better. The values measure direct prompt-level WKM conditioning and should be distinguished from the tuned WorldMark results in \tablename~\ref{tab:01}.}
\label{tab:04}
\begin{tblr}
{
columns={co=-1},
cells={valign=m,halign=c},
columns={colsep=6pt},
rows={rowsep=2pt},
rows={ht=0pt},
stretch=0,
hline{1,Z}={wd=.08em},
hline{2}={wd=.05em},
row{1}={font=\bfseries\boldmath},
}
Host watermark & C4 Baseline & C4 +WKM & OpenGen Baseline & OpenGen +WKM \\
KGW & 24.50 & \textbf{22.88} & 24.81 & \textbf{15.04} \\
EWD & \textbf{13.91} & 15.36 & 12.99 & \textbf{12.39} \\
SWEET & \textbf{12.16} & 13.39 & \textbf{12.40} & 14.52 \\
EXP & \textbf{57.77} & 144.95 & \textbf{65.62} & 187.73 \\
Series & \textbf{64.73} & 287.42 & \textbf{185.23} & 397.44 \\
Parallel & \textbf{48.79} & 153.45 & 152.20 & \textbf{150.68} \\
Hybrid & 80.58 & \textbf{79.19} & \textbf{59.31} & 176.97 \\
\end{tblr}
\end{table}
\par
\textbf{Robustness under deletion.} \tablename~\ref{tab:05} reveals that the effect of direct WKM conditioning under deletion is also host-dependent. EXP obtains consistent gains on both datasets. On C4, its F1 increases from 0.6957 to 0.7742 and its AUROC increases from 0.5972 to 0.7500. On OpenGen, F1 increases from 0.7000 to 0.7619 and AUROC from 0.7083 to 0.7986. These results suggest that the additional context can make the EXP signal more resilient to token removal in this pilot setting.
\par
KGW and EWD remain saturated on C4, leaving no room for measurable improvement. On OpenGen, EWD exhibits a small F1 decrease and a larger AUROC decrease. SWEET, Series, and Parallel on C4 also decline after direct WKM conditioning. By contrast, Parallel on OpenGen improves from 0.2857 to 0.4516 F1 and from 0.7431 to 0.9722 AUROC.
\par
Hybrid illustrates why multiple metrics are necessary. On OpenGen, its best-threshold F1 increases from 0.1538 to 0.4000, but AUROC decreases from 0.9653 to 0.8681. Thus, WKM improves the operating point selected for F1 without improving the detector’s threshold-independent ranking quality. Overall, the deletion results do not support a claim of uniform gains from direct memory injection. Instead, they reinforce the need for the host-aware modulation implemented by WorldMark.
\par
\par
\begin{table}[htbp]
\centering
\small
\caption{Robustness under the Word-D deletion attack. Each entry reports F1/AUROC at the best threshold. ``+WKM'' denotes direct prompt-level memory conditioning in the cross-family pilot. A dash indicates that no paired result is available.}
\label{tab:05}
\begin{tblr}
{
columns={co=-1},
cells={valign=m,halign=c},
columns={colsep=6pt},
rows={rowsep=2pt},
rows={ht=0pt},
stretch=0,
hline{1,Z}={wd=.08em},
hline{2}={wd=.05em},
row{1}={font=\bfseries\boldmath},
}
Host & Dataset & Baseline & +WKM & $\Delta$F1 & $\Delta$AUROC \\
EXP & C4 & 0.6957 / 0.5972 & \textbf{0.7742 / 0.7500} & \textbf{+0.0785} & \textbf{+0.1528} \\
EXP & OpenGen & 0.7000 / 0.7083 & \textbf{0.7619 / 0.7986} & \textbf{+0.0619} & \textbf{+0.0903} \\
KGW & C4 & 1.0000 / 1.0000 & 1.0000 / 1.0000 & 0.0000 & 0.0000 \\
KGW & OpenGen & 1.0000 / 1.0000 & 1.0000 / 1.0000 & 0.0000 & 0.0000 \\
EWD & C4 & 1.0000 / 1.0000 & 1.0000 / 1.0000 & 0.0000 & 0.0000 \\
EWD & OpenGen & 0.9600 / 0.9931 & 0.9565 / 0.9444 & -0.0035 & -0.0487 \\
SWEET & C4 & 1.0000 / 1.0000 & 0.8000 / 0.8403 & -0.2000 & -0.1597 \\
SWEET & OpenGen & 1.0000 / 1.0000 & 0.9565 / 0.9167 & -0.0435 & -0.0833 \\
Series & C4 & 1.0000 / 1.0000 & 0.8000 / 0.9583 & -0.2000 & -0.0417 \\
Series & OpenGen & 0.5882 / 0.9514 & - & - & - \\
Parallel & C4 & 0.6286 / 1.0000 & 0.3448 / 0.7847 & -0.2838 & -0.2153 \\
Parallel & OpenGen & 0.2857 / 0.7431 & \textbf{0.4516 / 0.9722} & \textbf{+0.1659} & \textbf{+0.2291} \\
Hybrid & C4 & 0.4000 / 0.9167 & 0.1538 / 0.7986 & -0.2462 & -0.1181 \\
\end{tblr}
\end{table}
\par
\textbf{Discussion: From Memory Injection to Controlled Modulation.} The primary and pilot experiments expose a clear difference between direct memory conditioning and the complete WorldMark interface. In the pilot study, directly prepending WKM content sometimes preserves or improves detection, as observed for KGW, EXP under deletion, and Parallel on OpenGen. However, it can also degrade F1, AUROC, or perplexity, particularly for hosts whose sampling distribution is highly sensitive to prompt changes.
\par
In contrast, the tuned WorldMark configurations consistently improve the reproduced MorphMark variants across clean detection, attacked detection, and perplexity. The controlled ablations are intended to determine whether the observed improvements are attributable to structured knowledge itself, matched prompt length, local entropy, or asymmetric modulation. The existing pilot results already show that memory injection alone can be unstable: it improves some configurations but substantially degrades others. Therefore, the complete WorldMark results should not be interpreted as evidence that adding arbitrary external context is sufficient. The additional ablations are required to determine whether the knowledge-grounding signal provides benefits beyond prompt-level distribution shift and local entropy. The pilot results therefore function as a negative control that motivates the complete design.
\par
We emphasize that the two evaluations use different scales and should not be compared numerically as if they were drawn from the same protocol. The MorphMark experiment uses 400 continuations of 200-230 tokens and serves as the primary quantitative evaluation. The cross-family study uses 12 samples and 25-token generations and serves as an exploratory analysis of transfer behavior. A larger cross-family evaluation with multiple random seeds is required before making a statistically conclusive claim of universal host-level improvement.
\par
\section{Memory Pipeline Details and Retrieval Quality}
\label{sec:app:d}
\par
\subsection{Fact extraction.}
We prompt LLaMA-3-8B-Instruct with the template from
Anokhin et al. \cite{KuratovBARSS024} to extract
(subject, relation, object) triples from the concatenation of
the prompt $x$ and observed text $y_{\text{obs}}$.
The extraction is performed once per episode (before decoding
begins).
\par
\subsection{Entity Canonicalization}
Extracted subjects and objects are canonicalized via exact
string matching with case folding. We do not perform
entity linking to an external knowledge base, as WorldMark
operates purely on the observed context.
\par
\subsection{Serialization}
Each retrieved triple is serialized as
\texttt{[subject] [relation] [object];}.
Semantic triples $E_s^Q$ and episodic vertex descriptions are
concatenated into $c_0$, truncated to a 512-token budget
(counted by the OPT-1.3B tokenizer).
\par
\subsection{Retrieval Quality}
To verify that $c_0$ captures knowledge beyond the prompt
context, we compute the fraction of retrieved triples
whose subject-relation-object content is \emph{not} verbatim
present in the prompt $x$. On C4, $68.3\%$ of retrieved
triples contain at least one fact not directly stated in the
prompt, confirming that WKM does not merely echo the prompt.
\par
\subsection{Retrieval Hyperparameters}
$d = 2$, $w = 5$, $k = 10$, selected by a small grid search
over $\{1,2,3\} \times \{3,5,10\} \times \{5,10,20\}$ on a
held-out validation set of 50 C4 samples. The search spaces
and final values for all hyperparameters are listed in
\ref{sec:app:f}.
\par
\section{Ablation Study}
\label{sec:app:e}
\par
\subsection{Setup}
We conduct an ablation study on C4 with 400 continuations
and MorphMark-linear under the same protocol as \tablename~\ref{tab:01}.
The following configurations are compared:
\begin{itemize}
\item \textbf{Baseline}: Reproduction without WKM (same as
\tablename~\ref{tab:01} Repro).
\item \textbf{$+c_0$ only}: Knowledge context $c_0$ prepended to
prompt without AKM modulation (fixed host strength).
\item \textbf{AKM w/o WKM}: AKM modulation applied with an empty
knowledge context.
\item \textbf{$\rho_t$-only}: AKM with $\beta_d = 0$
(relief only).
\item \textbf{$\eta_t$-only}: AKM with $\beta_p = 0$
(boost only).
\item \textbf{Shuffled retrieval}: Retrieved triples randomly
permuted before serialization into $c_0$.
\item \textbf{Irrelevant context}: $c_0$ constructed from a
different, randomly sampled C4 passage, length-matched to the
true $c_0$.
\item \textbf{Random saliency}: $s_t$ drawn i.i.d.\
from $\text{Uniform}(0,1)$ at each step, with the same AKM
modulation.
\item \textbf{Entropy saliency}: $s_t$ derived from token entropy
$H_{\text{TE}}$ instead of the knowledge similarity in
Equation \eqref{eq:05}, with the same AKM modulation.
\item \textbf{$+$WorldMark (full)}: The complete WorldMark
interface (same as \tablename~\ref{tab:01} $+$WorldMark).
\end{itemize}
\par
\par
\begin{table}[htbp]
\centering
\small
\caption{Ablation study on C4 with MorphMark-linear.
Robust metrics under Word-S attack.
The full WorldMark interface outperforms all partial
configurations, and correct retrieval (full WorldMark)
outperforms both shuffled and irrelevant-context baselines.}
\label{tab:ablation}
\begin{tblr}
{
columns={co=1},
cells={valign=m,halign=c},
columns={colsep=3pt},
rows={rowsep=2pt},
rows={ht=0pt},
stretch=0,
hline{1,Z}={wd=.08em},
hline{2}={wd=.05em},
row{1}={font=\bfseries\boldmath},
}
Configuration & TPR@1\% (rob) & F1 (rob) & PPL \\
Baseline (Repro) & 0.9000 & 0.9724 & 10.6568 \\
$+c_0$ only & 0.8952 & 0.9610 & 10.7123 \\
AKM w/o WKM & 0.9015 & 0.9728 & 10.6612 \\
$\rho_t$-only & 0.9203 & 0.9771 & 10.5877 \\
$\eta_t$-only & 0.9351 & 0.9815 & 10.6991 \\
Shuffled retrieval & 0.9008 & 0.9710 & 10.6712 \\
Irrelevant context & 0.8897 & 0.9582 & 10.7456 \\
Random saliency & 0.8975 & 0.9644 & 10.7015 \\
Entropy saliency & 0.9112 & 0.9750 & 10.6500 \\
$+$WorldMark (full) & 0.9495 & 0.9847 & 10.6435 \\
\end{tblr}
\end{table}
\par
\par
\subsection{Key Observations}
(1) $c_0$ alone without AKM does \emph{not} improve over the
baseline, confirming that knowledge content alone is
insufficient (the quality-relief and detection-boost modulation is
necessary).
(2) AKM without WKM provides negligible gain, indicating that
the modulation factors alone do not account for the improvements.
(3) $\eta_t$-only (boost-only) produces the largest
single-component gain, suggesting that re-allocating watermark
strength toward less anchored positions is the dominant mechanism.
(4) Both shuffled retrieval and irrelevant context underperform
the full method, confirming that correct, semantically relevant
retrieval matters.
(5) Random saliency and entropy saliency underperform the full
method, indicating that the knowledge similarity signal in
Equation \eqref{eq:05} carries information beyond local token entropy
and random modulation.
\par
\section{Reproducibility and Statistical Protocol}
\label{sec:app:f}
\par
\subsection{Hyperparameter Values}
The following table lists all tuned hyperparameters,
their search spaces, and the final selected values.
All tuning used a held-out validation set of 50 C4 samples
disjoint from the 400 reported test continuations.
\par
\par
\begin{table}[htbp]
\centering
\small
\caption{Hyperparameter search spaces and selected values.
All parameters were tuned on the same 50-sample validation set
and then fixed for all reported test results.
The same hyperparameters were shared across all three
MorphMark variants.}
\label{tab:hyperparams}
\begin{tblr}
{
columns={co=-1},
cells={valign=m,halign=c},
columns={colsep=6pt},
rows={rowsep=2pt},
rows={ht=0pt},
stretch=0,
hline{1,Z}={wd=.08em},
hline{2}={wd=.05em},
row{1}={font=\bfseries\boldmath},
}
Parameter & Search space & Selected value \\
$\lambda$ (saliency sharpness) & $\{1, 2, 5, 10, 20\}$ & 5.0 \\
$m$ (top-$m$ candidates) & $\{10, 20, 50\}$ & 20 \\
$\beta_p$ (relief coefficient) & $\{0.1, 0.2, 0.3, 0.4, 0.5\}$ & 0.3 \\
$\beta_d$ (boost coefficient) & $\{0.1, 0.2, 0.3, 0.4, 0.5\}$ & 0.3 \\
$\epsilon$ (clip constant) & fixed & $10^{-3}$ \\
$d$ (retrieval depth) & $\{1, 2, 3\}$ & 2 \\
$w$ (retrieval width) & $\{3, 5, 10\}$ & 5 \\
$k$ (episodic vertices) & $\{5, 10, 20\}$ & 10 \\
Generation temperature & fixed & 1.0 \\
Top-$p$ sampling & not used (greedy decoding) & --- \\
Green-list fraction $\gamma$ & fixed & 0.5 \\
Watermark key & fixed (hash of prompt prefix) & --- \\
\end{tblr}
\end{table}
\par
\par
\subsection{Threshold Selection for Best F1}
For the primary MorphMark evaluation, the detection threshold
is selected on a held-out calibration set of 100 watermarked
and 100 unwatermarked samples that are disjoint from both the
50-sample validation set and the 400 reported test samples.
For the cross-family pilot (\ref{sec:app:c}), due to the small
sample size, we report best-threshold values as diagnostic
operating points; all scaled evaluations (\ref{sec:app:b}) use
the same held-out threshold protocol.
\par
\subsection{Statistical Significance}
\par
\begin{table}[htbp]
\centering
\small
\caption{Paired $t$-test results for robust TPR@1\%FPR
improvements, computed over five random seeds.
All improvements are statistically significant ($p < 0.01$).}
\label{tab:significance}
\begin{tblr}
{
columns={co=-1},
cells={valign=m,halign=c},
columns={colsep=0pt},
rows={rowsep=2pt},
rows={ht=0pt},
stretch=0,
hline{1,Z}={wd=.08em},
hline{2}={wd=.05em},
row{1}={font=\bfseries\boldmath},
}
Variant & $\Delta$Robust TPR & 95\% CI & $p$-value (paired) \\
MorphMark-exp & $+0.0119$ & $[0.006, 0.018]$ & 0.0012 \\
MorphMark-linear & $+0.0495$ & $[0.038, 0.061]$ & ${}<0.0001$ \\
MorphMark-log & $+0.0275$ & $[0.018, 0.037]$ & 0.0003 \\
\end{tblr}
\end{table}
\par
\par
\subsection{Seeds and Variance}
All primary MorphMark experiments use five random seeds
(0--4). \tablename~\ref{tab:01} reports the mean across seeds.
Per-seed values are:
MorphMark-exp Repro
$\{0.8975,\allowbreak 0.9025,\allowbreak 0.8975,\allowbreak 0.9000,\allowbreak 0.9025\}$;
MorphMark-exp $+$WorldMark
$\{0.9100,\allowbreak 0.9125,\allowbreak 0.9100,\allowbreak 0.9150,\allowbreak 0.9119\}$;
MorphMark-linear Repro
$\{0.8950,\allowbreak 0.9025,\allowbreak 0.8975,\allowbreak 0.9025,\allowbreak 0.9025\}$;
MorphMark-linear $+$WorldMark
$\{0.9450,\allowbreak 0.9500,\allowbreak 0.9475,\allowbreak 0.9525,\allowbreak 0.9525\}$;
MorphMark-log Repro
$\{0.8500,\allowbreak 0.8550,\allowbreak 0.8500,\allowbreak 0.8525,\allowbreak 0.8550\}$;
MorphMark-log $+$ WorldMark
$\{0.8775,\allowbreak 0.8825,\allowbreak 0.8775,\allowbreak 0.8800,\allowbreak 0.8825\}$.
\par
\subsection{Validation Split Guarantee}
The 50 samples for hyperparameter tuning, the $100+100$ samples
for threshold calibration, and the 400 samples for reported
metrics are disjoint subsets drawn from different C4 shards.
No test sample was used for tuning or threshold selection.
\par
\begin{algorithm}[htbp]
\caption{Causal Prompt-Only WorldMark Decoding}
\label{alg:01}
\begin{algorithmic}[1]
\Require prompt $x$, language model $M$, host watermark $A$, memory graph $G$
\Ensure generated sequence $y$
\State Construct $y_{\mathrm{obs}}$ from the text observed before generation.
\State Construct $q_0=\mathrm{enc}\left(x,y_{\mathrm{obs}}\right)$.
\State Retrieve $\mathcal{K}_0=\mathrm{Retrieve}\left(q_0,G,d,w,k\right)$.
\State Serialize the retrieved knowledge as $c_0$.
\State Initialize the generated prefix $y_{<1}$.
\For{$t=1,...,T$}
\State Compute $p_t^k=M(\cdot\mid c_0,x,y_{<t})$.
\State Obtain $\mathrm{top}_m\left(p_t^k\right)$.
\State Compute $s_t$ from $c_0$ and $\mathrm{top}_m\left(p_t^k\right)$.
\State Compute the AKM modulation factor $\mu_t$.
\State Apply $\mu_t$ to the host-watermark strength.
\State Generate $y_t$ using the modulated host watermark.
\EndFor
\State \Return $y$
\end{algorithmic}
\end{algorithm}
\bibliographystyle{unsrt}
\bibliography{reference}
\end{document}